\documentclass[11pt,a4paper,american]{extarticle}
\usepackage[T1]{fontenc}
\usepackage[latin1]{inputenc}
\usepackage{amsmath}
\usepackage{amssymb}
\usepackage{setspace}
\usepackage{esint}
\makeatletter

\usepackage{babel}
\usepackage{bbm}
\usepackage{hyperref}
\usepackage{authblk}
\usepackage{tikz}
\usetikzlibrary{matrix}
\usepackage{cancel}
\date{}
\allowdisplaybreaks
\numberwithin{equation}{section}

\author[1]{Thales Azevedo\thanks{thales.azevedo@physics.uu.se}}
\author[2]{Renann Lipinski Jusinskas\thanks{renannlj@fzu.cz}}
\affil[1]{Department of Physics and Astronomy, Uppsala University \authorcr Box 516, 751 20,  Uppsala - Sweden\authorcr \ }
\affil[2]{Institute of Physics AS CR \authorcr  Na Slovance 2, 182 21, Prague - Czech Republic}

\makeatother

\usepackage{babel}
\begin{document}

\title{Connecting the ambitwistor and the sectorized heterotic strings }
\maketitle
\begin{abstract}
The sectorized description of the (chiral) heterotic string using
pure spinors has been misleadingly viewed as an infinite tension string.
One evidence for this fact comes from the tree level 3-point graviton
amplitude, which we show to contain the usual Einstein term plus a
higher curvature contribution. After reintroducing a dimensionful
parameter $\ell$ in the theory, we demonstrate that the heterotic
model is in fact two-fold, depending on the choice of the supersymmetric
sector, and that the spectrum also contains one massive (open string
like) multiplet. By taking the limit $\ell\to\infty$, we finally
show that the ambitwistor string is recovered, reproducing the unexpected
heterotic state in Mason and Skinner's RNS description. \tableofcontents{}
\end{abstract}

\section{Introduction\label{sec:Introduction}}

After the discovery of the remarkable CHY formulae by Cachazo, He
and Yuan \cite{Cachazo:2013hca}, which compute N-point tree-level
amplitudes of massless states through an integration over points on
the Riemann sphere, a natural question was posed: Is there a deeper
mathematical framework underpinning them?

It did not take long before an elegant answer to that question was
proposed. Mason and Skinner presented in \cite{Mason:2013sva} the
so-called ambitwistor string theory, consisting of a chiral worldsheet
model whose tree-level correlation functions give precisely the CHY
formulae. Shortly after, Berkovits introduced a manifestly supersymmetric
version of the ambitwistor strings \cite{Berkovits:2013xba}, described
by a chiral form of the usual pure spinor string. The amplitudes computed
using this formalism were later shown to give rise to a supersymmetric
version of the CHY amplitudes \cite{Gomez:2013wza}.

In the heterotic case, however, both models are not totally satisfying.
When restricted to single-trace contributions, the super-Yang-Mills
sector gives the expected result for the amplitudes. On the other
hand, the supergravity sector failed to reproduce the usual Einstein
term. Moreover it seemed impossible in the pure spinor description
to find a supergravity vertex operator with the correct properties.
This is related to the fact that Berkovits' proposal for the BRST
charge does not contain the particle-like Hamiltonian, $H=-\frac{1}{2}P^{m}P_{m}$,
which then fails to encode the expected gauge transformations in terms
of BRST exact states.

This issue did not receive much attention until one of the authors,
inspired by the works of Chand\'ia and Vallilo \cite{Chandia-Vallilo},
developed a new pure-spinor model, referred to as the sectorized heterotic
string in the present paper, in which the BRST charge is modified
with respect to the original proposal by Berkovits and the description
of supergravity states was made possible \cite{Jusinskas:2016qjd}.
As a consistency check for the new BRST charge, the sectorized string
was later coupled to a generic heterotic background \cite{Azevedo:2016zod}
and we were able to show that requiring (classical) nilpotency of
the BRST charge is equivalent to imposing the classical constraints
on the background superfields, which had been previously shown by
Berkovits and Howe to imply the supergravity and super Yang-Mills
equations of motion \cite{Berkovits:2001ue}.

Providing a sensible vertex operator for the states in the supergravity
sector was an important achievement of the sectorized model\footnote{More precisely, while the unintegrated vertex operators are known,
the integrated ones are still missing in the sectorized models. Although
this is not the focus of this work, we briefly discuss this topic
in section \ref{sec:conclusion}.}. One would then be able to use that vertex operator to compute 3-point,
tree-level correlation function of gravitons, and hopefully to reproduce
the usual amplitude coming from Einstein gravity. However, that is
not the case, as we will later show. In fact, the 3-point amplitude
includes the usual terms coming from Einstein gravity but it also
contains terms of order four in the momenta, which resemble a Gauss-Bonnet
theory. 

The presence of a higher-order term in the 3-point amplitude motivated
us to reintroduce a dimensionful parameter in the theory and then
look for massive states. We found that in addition to the massless
states the BRST cohomology contains a single massive level, which
is reminiscent of the first massive level of the open superstring.
We investigated these extra states in detail, as well as their relation
to the original heterotic ambitwistor string. In particular, there
is a massive 3-form which reduces in the infinite length (tensionless)
limit to the unexpected massless 3-form originally encountered by
Mason and Skinner. This can be explicitly shown in an $SO(8)$-covariant
form via a Del Giudice-Di Vecchia-Fubini (DDF) construction of the
vertex operator.

In \cite{Siegel:2015axg}, Siegel already observed that ambitwistor
strings could be viewed as tensionless string-like models. The connection
between the sectorized model and the ambitwistor string as its infinite
length limit is similar to the one briefly considered by Casali and
Tourkine in \cite{Casali:2016atr} for the bosonic string, which in
turn is closely related to the work of Huang, Siegel and Yuan in \cite{Huang:2016bdd}.
Indeed, there are many similarities between our results and those
obtained by Huang \emph{et al}. In particular, the 3-point graviton
amplitude we compute in the present paper agrees with their results
for the chiral heterotic string, including the two different cases
depending on which sector is ``flipped''. In our case, this corresponds
to the two inequivalent ways of sectorizing the string, which we call
A and B models. As usual, the pure spinor approach is more covenient
in the supersymmetric analysis.

As already mentioned, we give compelling evidence that the tensionless
limit of the sectorized string is equivalent to the ambitwistor string.
But more surprisingly, our results also seem to be connected to the
twistor string theory studied by Berkovits and Witten \cite{Witten:2003nn,Berkovits:2004jj}.
In \cite{Azevedo:2017lkz} a formula for the heterotic ambitwistor
amplitude of $n$ gravitons was found which reduces to that of Berkovits-Witten
in the four-dimensional MHV case. We believe this subject deserves
more investigation, and expect to address it in future work\footnote{We are thankful to Henrik Johansson for bringing this to our attention.}.

This paper is organized as follows. In section \ref{sec:Amodel},
after reviewing the sectorized model, we present the tree level 3-point
graviton amplitude with its different contributions, motivating the
introduction of a length parameter $\ell$ and a more thorough study
of the cohomology. In section \ref{sec:Bmodel}, we propose an alternative
definition of the heterotic sectorized string, demonstrating that
the physical spectrum contains also one massive (open string like)
multiplet. Section \ref{sec:ambitwistor} is dedicated to the analysis
of the $\ell\to\infty$ limit and a new proposal for the heterotic
ambitwistor BRST charge in the pure spinor formalism and its cohomology.
We summarize our results and discuss open problems in section \ref{sec:conclusion}.
Appendix \ref{sec:OPE} lists the relevant OPE's used in the paper.

\section{The sectorized model and the dimensionful parameter\label{sec:Amodel}}

The action for the heterotic sectorized string in the pure spinor
formalism is given by
\begin{equation}
S=\int d^{2}z\{P_{m}\bar{\partial}X^{m}+p_{\alpha}\bar{\partial}\theta^{\alpha}+w_{\alpha}\bar{\partial}\lambda^{\alpha}+\bar{b}\bar{\partial}\bar{c}+\mathcal{L}_{C}\},\label{eq:heteroticaction}
\end{equation}
where $(X^{m},\theta^{\alpha})$ denote the target space supercoordinates
of the string and $(P_{m},p_{\alpha})$ their conjugate momenta, with
$m=0,\ldots,9$ and $\alpha=1,\ldots,16$; $\lambda^{\alpha}$ is
the pure spinor satisfying $(\lambda\gamma^{m}\lambda)=0$ and $w_{\alpha}$
its conjugate; $(\bar{b},\bar{c})$ is the usual Virasoro ghost pair.
The gamma matrices, $\gamma_{\alpha\beta}^{m}$ and $\gamma^{m\alpha\beta}$,
satisfy $\{\gamma^{m},\gamma^{n}\}=2\eta^{mn}$, where $\eta^{mn}$
is the $SO(9,1)$ metric. The Lagrangian $\mathcal{L}_{C}$ describes
the $SO(32)$ or $E(8)\times E(8)$ gauge sector with energy-momentum
$T_{C}$ and central charge $16$, and the associated current algebra
is realized by the (holomorphic) generators $J^{I}$, with $I$ denoting
the adjoint representation of the gauge group.

The fundamental feature of this chiral model is its interpretation
in terms of two sectors, $(+)$ and $(-)$, that resemble the usual
left and right movers of the superstring \cite{Jusinskas:2016qjd}.
There are two decoupled sets of operators, $\mathcal{O}_{\pm}$, organized
as\footnote{There are in fact two inequivalent ways of spliting the worldsheet
operators in two decoupled sets, as explained in section \ref{sec:Bmodel}.
Here we review the choice made in previous works \cite{Jusinskas:2016qjd,Azevedo:2016zod},
referred to in the present paper as the A model.}\begin{subequations}
\begin{eqnarray}
\mathcal{O}_{+} & = & \{P_{m}^{+},\bar{b},\bar{c},J^{I}\},\\
\mathcal{O}_{-} & = & \{P_{m}^{-},d_{\alpha},\theta^{\alpha},w_{\alpha},\lambda^{\alpha}\},
\end{eqnarray}
\end{subequations}with\begin{subequations}\label{eq:sectorOP}
\begin{eqnarray}
P_{m}^{+} & = & P_{m}+\partial X_{m},\label{eq:P+P+OPE}\\
P_{m}^{-} & = & P_{m}-\partial X_{m}-(\theta\gamma_{m}\partial\theta),\\
d_{\alpha} & = & p_{\alpha}-\frac{1}{2}(\gamma^{m}\theta)_{\alpha}P_{m}^{-}-\frac{1}{4}(\theta\gamma^{m}\partial\theta)(\gamma_{m}\theta)_{\alpha}.
\end{eqnarray}
\end{subequations}The $\mathcal{N}=1$ supersymmetry of the heterotic
string is compatible with the sectorized formulation and all the operators
displayed in \eqref{eq:sectorOP} are invariant under the supersymmetry
charge
\begin{equation}
q_{\alpha}=\oint\{p_{\alpha}+\frac{1}{2}P_{m}^{-}(\gamma^{m}\theta)_{\alpha}+\frac{5}{12}(\theta\gamma^{m}\partial\theta)(\gamma_{m}\theta)_{\alpha}\}.
\end{equation}

The two sectors have characteristic (pseudo) energy-momentum tensors
defined by\begin{subequations}\label{eq:T+T-het}
\begin{eqnarray}
T_{-} & \equiv & \frac{1}{4}\eta^{mn}P_{m}^{-}P_{n}^{-}-d_{\alpha}\partial\theta^{\alpha}-w_{\alpha}\partial\lambda^{\alpha},\\
T_{+} & \equiv & -\frac{1}{4}\eta^{mn}P_{m}^{+}P_{n}^{+}+T_{C}-\bar{b}\partial\bar{c}-\partial(\bar{b}\bar{c}),
\end{eqnarray}
\end{subequations}which combine to form the energy-momentum tensor
associated to the action \eqref{eq:heteroticaction},
\begin{eqnarray}
T & = & T_{+}+T_{-},\nonumber \\
 & = & -P_{m}\partial X^{m}-p_{\alpha}\partial\theta^{\alpha}-w_{\alpha}\partial\lambda^{\alpha}-2\bar{b}\partial\bar{c}-\bar{c}\partial\bar{b}+T_{C}.
\end{eqnarray}

The BRST charge of the model is defined to be
\begin{equation}
Q=\oint\{\lambda^{\alpha}d_{\alpha}+\bar{c}T_{+}-\bar{b}\bar{c}\partial\bar{c}\},\label{eq:QPSsect}
\end{equation}
and naturally incorporates the sector splitting. Furthermore, its
cohomology consistently describes the expected heterotic massless
spectrum. In addition to the super Yang-Mills states, encoded by the
superfield $A_{\alpha}^{I}$ in the vertex operator $U_{SYM}$,\begin{subequations}\label{eq:SYMvertex}
\begin{eqnarray}
U_{SYM} & = & \lambda^{\alpha}\bar{c}A_{\alpha}^{I}J_{I},\\
\gamma_{mnpqr}^{\alpha\beta}D_{\alpha}A_{\beta}^{I} & = & 0,\\
\delta_{\Sigma}A_{\alpha}^{I} & = & D_{\alpha}\Sigma^{I},\label{eq:gaugeSYM}
\end{eqnarray}
\end{subequations}the cohomology of $Q$ contains also the $\mathcal{N}=1$
supergravity states, encoded by the superfield $A_{\alpha}^{m}$ in
the vertex operator $U_{SG}$,\begin{subequations}\label{eq:SUGRAvertex}
\begin{eqnarray}
U_{SG} & = & \lambda^{\alpha}\bar{c}A_{\alpha}^{m}P_{m}^{+},\label{eq:SGvertex}\\
\gamma_{mnpqr}^{\alpha\beta}D_{\alpha}A_{\beta}^{s} & = & 0,\\
\partial^{n}\partial_{n}A_{\alpha}^{m}-\partial^{m}\partial_{n}A_{\alpha}^{n} & = & 0,\\
\delta_{\Sigma}A_{\alpha}^{m} & = & D_{\alpha}\Sigma^{m}+\partial^{m}\Sigma_{\alpha}.\label{eq:gaugeSG}
\end{eqnarray}
\end{subequations}The superfields $\Sigma^{I}$, $\Sigma_{\alpha}$
and $\Sigma^{m}$ are the gauge parameters of the transformations
\eqref{eq:gaugeSYM} and \eqref{eq:gaugeSG}. More details can be
found in \cite{Jusinskas:2016qjd}.

It is worthwhile to point out that the supergravity sector does not
have a clear description in the ambitwistor heterotic string \cite{Mason:2013sva,Berkovits:2013xba}.
Providing a sensible vertex operator for the supergravity states was
an important achievement of the sectorized model. Going further, we
were also able to show in \cite{Azevedo:2016zod} that the model imposes
the correct (on-shell) constraints when coupled to a generic heterotic
background. As another test for the sectorized model, we can compute
the 3-point, tree-level correlation function of gravitons using the
vertex operator \eqref{eq:SGvertex}, as we show in the following
subsection.

\subsection{3-point tree-level amplitude for gravitons}

It is clear that the vertex \eqref{eq:SGvertex} has a very similar
structure to the ordinary vertex in the full string. Thus, thinking
of the sectorized string as an infinite tension limit of the usual
string, one would expect the 3-point graviton amplitude to give rise
to the usual expression coming from Einstein gravity. However, that
is not the case, as seen from the computation we now sketch.

The unintegrated vertex operator for the graviton labeled by $j$
($j\in\{1,2,3\}$) is given by 
\begin{equation}
U_{(j)}=\bar{c}\lambda^{\alpha}A_{(j)}{}_{\alpha}^{m}(\theta)P_{m}^{+}e^{ik_{(j)}\cdot X},
\end{equation}
and $A_{(j)}{}_{\alpha}^{m}(\theta)$ can be gauged to the following
expansion, 
\begin{equation}
A_{(j)}{}_{\alpha}^{m}(\theta)=\frac{1}{2}(\gamma_{n}\theta)_{\alpha}\epsilon_{(j)}^{mn}-\frac{i}{16}(\gamma^{q}\theta)_{\alpha}(\theta\gamma_{npq}\theta)k_{(j)}^{n}\epsilon_{(j)}^{mp}+\mathcal{O}(\theta^{5}),
\end{equation}
where $\epsilon_{(j)}^{mn}$ is a symmetric-traceless polarization
tensor such that $\eta_{mn}k_{(j)}^{n}\epsilon_{(j)}^{mp}=0$.

The 3-point scattering amplitude at tree level is given by the usual
correlator of three vertex operators inserted at arbitrary points
$z_{j}$ on the 2-sphere: 
\begin{equation}
\langle U_{(1)}(z_{1})U_{(2)}(z_{2})U_{(3)}(z_{3})\rangle=\langle\bar{c}(z_{1})\bar{c}(z_{2})\bar{c}(z_{3})\rangle\langle\lambda^{\alpha}A_{(1)}{}_{\alpha}^{m}(\theta)\lambda^{\beta}A_{(2)}{}_{\beta}^{n}(\theta)\lambda^{\gamma}A_{(3)}{}_{\gamma}^{p}(\theta)\rangle C_{XP},\label{correlators}
\end{equation}
where $C_{XP}$ is the correlator involving $X$ and $P^{+}$ and
we have separated the independent contributions, to which we now turn.

The $\bar{c}$-ghost correlator is the same as in bosonic string theory
and is known to give (up to an overall factor) 
\begin{equation}
\langle\bar{c}(z_{1})\bar{c}(z_{2})\bar{c}(z_{3})\rangle=z_{12}z_{23}z_{31}\qquad(z_{ij}:=z_{i}-z_{j}).
\end{equation}
The $\lambda$ and $\theta$ zero-mode integration is performed via
the usual BRST-invariant measure factor $\langle(\lambda\gamma^{m}\theta)(\lambda\gamma^{n}\theta)(\lambda\gamma^{p}\theta)(\theta\gamma_{mnp}\theta)\rangle=1$,
such that the second correlator in \eqref{correlators} reduces to
(again up to normalization) 
\begin{equation}
i(k_{(3)}\cdot\epsilon_{(1)})^{m}\eta_{qr}\epsilon_{(2)}^{nq}\epsilon_{(3)}^{pr}+i(k_{(1)}\cdot\epsilon_{(2)})^{n}\eta_{qr}\epsilon_{(3)}^{pq}\epsilon_{(1)}^{mr}+i(k_{(2)}\cdot\epsilon_{(3)})^{p}\eta_{qr}\epsilon_{(1)}^{mq}\epsilon_{(2)}^{nr}.
\end{equation}
Finally, there are two contributions to $C_{XP}$. The first one involves
Wick contractions between $X$ and $P^{+}$ and also between two $P^{+}$'s
(note that $P^{+}$ has a nontrivial OPE with itself due to the mixing
of $P^{m}$ and $\partial X^{m}$, \emph{cf}. equation \eqref{eq:P+P+OPE}).
The second one involves only contractions between $X$ and $P^{+}$.
Recall that there are no contractions between the exponentials since
the $XX$ OPE is trivial. The first set of contractions gives 
\begin{multline}
2i[z_{12}^{-2}\eta_{mn}(z_{31}^{-1}k_{(1)p}+z_{32}^{-1}k_{(2)p})+z_{23}^{-2}\eta_{np}(z_{12}^{-1}k_{(2)m}+z_{13}^{-1}k_{(3)m})+z_{31}^{-2}\eta_{mp}(z_{23}^{-1}k_{(3)n}+z_{21}^{-1}k_{(1)n})]=\\
=-2i(z_{12}z_{23}z_{31})^{-1}(\eta_{mn}k_{(1)p}+\eta_{np}k_{(2)m}+\eta_{mp}k_{(3)n}),
\end{multline}
where we have used momentum conservation and transversality of the
polarization tensors, for example, 
\begin{equation}
z_{31}^{-1}k_{(1)p}+z_{32}^{-1}k_{(2)p}=(z_{31}^{-1}-z_{32}^{-1})k_{(1)p}=\frac{z_{12}}{z_{31}z_{32}}k_{(1)p}.
\end{equation}

Combining these results and using the shorthand notation $\epsilon_{(i)}^{mn}\equiv\epsilon_{(i)}^{m}\epsilon_{(i)}^{n}$,
the first contribution to the amplitude can be expressed as 
\begin{equation}
\mathcal{M}_{1}=2[(\epsilon_{(1)}\cdot\epsilon_{(2)})(\epsilon_{(3)}\cdot k_{(1)})+(\epsilon_{(2)}\cdot\epsilon_{(3)})(\epsilon_{(1)}\cdot k_{(2)})+(\epsilon_{(3)}\cdot\epsilon_{(1)})(\epsilon_{(2)}\cdot k_{(3)})]^{2},
\end{equation}
just like the usual term coming from Einstein gravity.

To conclude the computation of this amplitude, we need to find the
second contribution in $C_{XP}$. There are two independent ways of
contracting the $P^{+}$'s with the $X$'s in the exponentials, and
the result is
\begin{equation}
i(z_{12}z_{23}z_{31})^{-1}(k_{(2)m}k_{(3)n}k_{(1)p}-k_{(3)m}k_{(1)n}k_{(2)p}).
\end{equation}
Thus the second contribution to the amplitude is given by
\begin{multline}
\mathcal{M}_{2}=2[(\epsilon_{(1)}\cdot\epsilon_{(2)})(\epsilon_{(3)}\cdot k_{(1)})+(\epsilon_{(2)}\cdot\epsilon_{(3)})(\epsilon_{(1)}\cdot k_{(2)})+(\epsilon_{(3)}\cdot\epsilon_{(1)})(\epsilon_{(2)}\cdot k_{(3)})]\times\\
\times(\epsilon_{(1)}\cdot k_{(2)})(\epsilon_{(2)}\cdot k_{(3)})(\epsilon_{(3)}\cdot k_{(1)}),\label{eq:GBamplitude}
\end{multline}
which can be interpreted in terms of a Gauss-Bonnet theory. More precisely,
any combination of the curvature-squared terms in a Lagrangian would
give rise to \eqref{eq:GBamplitude}, since the Ricci tensor vanishes
on-shell. Nonetheless, the Gauss-Bonnet choice renders the theory
ghost-free (see, for example, \cite{Zwiebach:1985uq}).

Note that $\mathcal{M}_{1}$ and $\mathcal{M}_{2}$ have different
powers in the gravitons' momenta, an unexpected feature of a model
thought to be dimensionless. In what follows we will discuss this
in detail.

\subsection{Dimensionful parameter\label{sec:parameter}}

A more careful look at the model raises some questions about the role
played by dimensionality. In this direction, the operators defined
in \eqref{eq:sectorOP} lack a dimensionful parameter, which in the
usual string would be represented by $\alpha'$.

The fundamental observation here is that the sectorized string was
originally thought of as an alternative formulation of the ambitwistor
string, which in turn is viewed as an infinite tension string ($\alpha'\to0$).
In order to avoid confusion, we will introduce a parameter of length
dimension denoted by $\ell$ and later on we will discuss its relation
to $\alpha'$.

To introduce $\ell$ consistently, we will take advantage of a global
scaling symmetry of the chiral action \eqref{eq:heteroticaction}
and rescale the worldsheet fields as
\begin{equation}
\begin{array}{cccc}
X^{m}\to\ell^{-1}X^{m}, & \theta^{\alpha}\to\ell^{-\frac{1}{2}}\theta^{\alpha}, & \lambda^{\alpha}\to\ell^{-\frac{1}{2}}\lambda^{\alpha}, & \bar{c}\to\ell^{-2}\bar{c},\\
P_{m}\to\ell^{+1}P_{m}, & p_{\alpha}\to\ell^{+\frac{1}{2}}p_{\alpha}, & w_{\alpha}\to\ell^{+\frac{1}{2}}w_{\alpha}, & \bar{b}\to\ell^{+2}\bar{b}.
\end{array}\label{eq:scale}
\end{equation}
The Lagrangian of the gauge sector, $\mathcal{L}_{C}$, will not change
and $J_{I}\to J_{I}$. These particular weights were chosen in such
a way that the BRST charge \eqref{eq:QPSsect} is finite in the limit
$\ell\to\infty$. Now we are able to dimensionally balance the supersymmetric
invariants defined before:\begin{subequations}\label{eq:heteroticsusyinvariants}
\begin{eqnarray}
d_{\alpha} & = & p_{\alpha}-\frac{1}{2}P_{m}(\gamma^{m}\theta)_{\alpha}+\frac{1}{2\ell^{2}}\Pi^{m}(\gamma_{m}\theta)_{\alpha},\label{eq:dell}\\
P_{m}^{-} & = & P_{m}-\frac{1}{\ell^{2}}\partial X_{m}-\frac{1}{\ell^{2}}(\theta\gamma_{m}\partial\theta),\\
P_{m}^{+} & = & P_{m}+\frac{1}{\ell^{2}}\partial X_{m}.\\
\Pi^{m} & = & \partial X^{m}+\frac{1}{2}(\theta\gamma^{m}\partial\theta),\label{eq:susymomentumheterotic}\\
T_{-} & \equiv & \frac{1}{4}\eta^{mn}P_{m}^{-}P_{n}^{-}-\frac{1}{\ell^{2}}(d_{\alpha}\partial\theta^{\alpha}+w_{\alpha}\partial\lambda^{\alpha}),\label{eq:T-het-ell}\\
T_{+} & \equiv & -\frac{1}{4}\eta^{mn}P_{m}^{+}P_{n}^{+}-\frac{1}{\ell^{2}}(2\bar{b}\partial\bar{c}-\bar{c}\partial\bar{b}-T_{C}).\label{eq:T+het-ell}
\end{eqnarray}
\end{subequations}Aside from $\Pi_{m}=\frac{\ell^{2}}{2}(P_{m}^{+}-P_{m}^{-})$,
the operators above correspond to those displayed at the beginning
of this section. Note that the energy-momentum tensor is given by
$T=\ell^{2}(T_{+}+T_{-})$ while the generalized particle like Hamiltonian
is $\mathcal{H}=(T_{+}-T_{-})$,
\begin{eqnarray}
\mathcal{H} & = & -\frac{1}{4}\eta^{mn}P_{m}^{+}P_{n}^{+}-\frac{1}{4}\eta^{mn}P_{m}^{-}P_{n}^{-}\nonumber \\
 &  & +\frac{1}{\ell^{2}}[T_{C}-\bar{b}\partial\bar{c}-\partial(\bar{b}\bar{c})+d_{\alpha}\partial\theta^{\alpha}+w_{\alpha}\partial\lambda^{\alpha}],
\end{eqnarray}
which is supersymmetric and BRST-exact \cite{Jusinskas:2016qjd}.

The BRST charge has the same form as before and can be organized as\begin{subequations}\label{eq:BRSTscaleA}
\begin{eqnarray}
Q & = & Q_{\lambda}+Q_{+},\\
Q_{\lambda} & \equiv & \oint\,\lambda^{\alpha}d_{\alpha},\\
Q_{+} & \equiv & \oint\,\bar{c}(T_{+}+\frac{1}{\ell^{2}}\bar{b}\partial\bar{c}),
\end{eqnarray}
\end{subequations}\emph{cf.} the operators displayed in \eqref{eq:heteroticsusyinvariants}.

After reintroducing this ``hidden'' parameter $\ell$ back into
the theory, it is reasonable to take a more careful look at the cohomology
of \eqref{eq:BRSTscaleA}. Considering only eigenstates of momentum,
it is possible to show that the massless vertex operator with ghost
number two has the general form
\begin{eqnarray}
U & = & \partial\bar{c}\lambda^{\alpha}A_{\alpha}+(P_{m}^{+},\bar{c}\lambda^{\alpha}A_{\alpha}^{m})+\bar{c}\lambda^{\alpha}J_{I}A_{\alpha}^{I}\nonumber \\
 &  & +(P_{m}^{+},\bar{c}\partial\bar{c}A^{m})+\lambda^{\alpha}\lambda^{\beta}A_{\alpha\beta}+\frac{1}{\ell^{2}}\bar{c}\partial^{2}\bar{c}A,\label{eq:masslessvertex}
\end{eqnarray}
with ordering prescription defined as
\begin{equation}
(\mathcal{O}_{1},\mathcal{O}_{2})(y)\equiv\frac{1}{2\pi i}\oint\frac{dz}{z-y}\mathcal{O}_{1}(z)\mathcal{O}_{2}(y).\label{eq:ordering}
\end{equation}
The superfields $A$, $A_{\alpha}$, $A^{m}$, $A_{\alpha}^{m}$,
$A_{\alpha}^{I}$ and $A_{\alpha\beta}$ have gauge transformations
given by\begin{subequations}
\begin{eqnarray}
\delta A_{\alpha} & = & D_{\alpha}\Omega,\label{eq:gaugeSMax1}\\
\delta A^{m} & = & -\frac{1}{2}\partial^{m}\Omega,\label{eq:gaugeSMax2}\\
\delta A_{\alpha\beta} & = & \frac{1}{2}D_{\alpha}\Omega_{\beta}+\frac{1}{2}D_{\beta}\Omega_{\alpha},\label{eq:gaugeantifield}\\
\delta A_{\alpha}^{m} & = & \frac{1}{2}\partial^{m}\Omega_{\alpha}-D_{\alpha}\Omega^{m},\\
\delta A_{\alpha}^{I} & = & D_{\alpha}\Omega^{I},\\
\delta A & = & -\Omega+\frac{1}{2}\partial_{m}\Omega^{m},
\end{eqnarray}
\end{subequations}and BRST-closedness of $U$ implies the following
equations of motion:\begin{subequations}
\begin{eqnarray}
\lambda^{\alpha}\lambda^{\beta}D_{\beta}A_{\alpha} & = & 0,\label{eq:SMax1}\\
D_{\alpha}A^{m}+\frac{1}{2}\partial^{m}A_{\alpha} & = & 0.\label{eq:SMax2}\\
D_{\alpha}A+A_{\alpha}+\frac{1}{2}\partial_{m}A_{\alpha}^{m} & = & 0,\label{eq:sugratrans}\\
\frac{1}{2\ell^{2}}\partial_{m}A^{m} & = & 0,\\
\lambda^{\alpha}\lambda^{\beta}\lambda^{\gamma}D_{\gamma}A_{\alpha\beta} & = & 0,\label{eq:antifield}\\
\lambda^{\alpha}\lambda^{\beta}D_{\beta}A_{\alpha}^{m}-\frac{1}{2}\lambda^{\alpha}\lambda^{\beta}\partial^{m}A_{\alpha\beta} & = & 0,\label{eq:sugra}\\
\lambda^{\alpha}\lambda^{\beta}D_{\beta}A_{\alpha}^{I} & = & 0.
\end{eqnarray}
\end{subequations}

The super Maxwell field is a solution of \eqref{eq:SMax1}. However,
equation \eqref{eq:SMax2} implies that $A_{\alpha}\neq0$ and $A^{m}\neq0$
are pure gauge, \emph{cf}. \eqref{eq:gaugeSMax1} and \eqref{eq:gaugeSMax2}.
The same argument holds for $A_{\alpha\beta}$. Equation \eqref{eq:antifield}
has a known solution, namely the antifields of super Maxwell \cite{Berkovits:2001rb}.
On the other hand, if $A_{\alpha\beta}$ is not zero, equation \eqref{eq:sugra}
implies it is pure gauge, \emph{cf}. equation \eqref{eq:gaugeantifield}.
Therefore $A_{\alpha}=A^{m}=A_{\alpha\beta}=0$. Equation \eqref{eq:sugratrans}
states that the longitudinal part of $A_{\alpha}^{m}$ is also pure
gauge, so we can use $\partial_{m}\Omega^{m}$ to fix $A=0$ and make
$A_{\alpha}^{m}$ transversal. In the end, we are left with the known
massless content given in \eqref{eq:SYMvertex} and \eqref{eq:SUGRAvertex},
respectively super Yang-Mills and $\mathcal{N}=1$ supergravity. Note
that the physical states have ghost number one with respect to each
sector, $(+)$ and $(-)$, but never ghost number two within the same
sector.

It is now natural to question whether the existence of massive states
has been so far ignored. For massive solutions, it is easy to show
that the vertex \eqref{eq:masslessvertex} is BRST-exact. However,
there is one possible vertex construction that was not considered
before,
\begin{equation}
U=\bar{c}U_{-},\label{eq:vertex-open}
\end{equation}
where $U_{-}$ is a ghost number one, conformal weight $+1$ operator
composed out of superfields combined with currents from the $(-)$
sector. Computing the commutator of \eqref{eq:vertex-open} with the
BRST charge, one obtains
\begin{equation}
[Q,U]=(\frac{1}{\ell^{2}}+\frac{\Box}{4})\bar{c}\partial\bar{c}U_{-}-\bar{c}\{Q_{\lambda},U_{-}\}.\label{eq:massive-tachyon}
\end{equation}
The two terms on the right hand side of the above equation are independent
and must vanish separately. The vanishing of the first one implies
the mass-shell condition $M^{2}=-\frac{4}{\ell^{2}}$, \emph{i.e.}
a tachyon. The vanishing of the second term implies that $U_{-}$
is in the cohomology of $Q_{\lambda}$, the usual pure spinor BRST
charge. These two conditions are incompatible simply because the cohomology
of $Q_{\lambda}$ is supersymmetric, ensuring the absence of tachyonic
states.

Therefore, a massive vertex like \eqref{eq:vertex-open} cannot possibly
be in the cohomology of the BRST charge \eqref{eq:BRSTscaleA}. However,
as we will present in the next section, there is an alternative formulation
of the heterotic sectorized string that indeed contains a massive
solution.

\section{Alternative formulation\label{sec:Bmodel}}

By examining the structure of the BRST transformation \eqref{eq:massive-tachyon},
we can directly identify the origin of the wave operator, $\Box$.
It comes from the piece $\bar{c}T_{+}$ of the BRST charge. Had
it been replaced by $\bar{c}T_{-}$, the wave operator would
change sign, potentially leading to a massive state in the cohomology.
At this point, we observe an ambiguity in defining the $(+)$ and
$(-)$ sectors of the heterotic model. Notice that instead of the
operator set \eqref{eq:heteroticsusyinvariants}, we can define\begin{subequations}\label{eq:susyinvariantsB}
\begin{eqnarray}
\hat{d}_{\alpha} & = & p_{\alpha}-\frac{1}{2}P_{m}(\gamma^{m}\theta)_{\alpha}-\frac{1}{2\ell^{2}}\Pi^{m}(\gamma_{m}\theta)_{\alpha},\\
\hat{P}_{m}^{+} & = & P_{m}+\frac{1}{\ell^{2}}\partial X_{m}+\frac{1}{\ell^{2}}(\theta\gamma_{m}\partial\theta),\\
\hat{P}_{m}^{-} & = & P_{m}-\frac{1}{\ell^{2}}\partial X_{m},
\end{eqnarray}
\end{subequations}and, as the characteristic energy-momentum tensors
of each sector,\begin{subequations}\label{eq:T+T-het-ell-massive}
\begin{eqnarray}
\hat{T}_{+} & \equiv & -\frac{1}{4}\eta^{mn}\hat{P}_{m}^{+}\hat{P}_{n}^{+}-\frac{1}{\ell^{2}}(\hat{d}_{\alpha}\partial\theta^{\alpha}+w_{\alpha}\partial\lambda^{\alpha}),\\
\hat{T}_{-} & \equiv & \frac{1}{4}\eta^{mn}\hat{P}_{m}^{-}\hat{P}_{n}^{-}-\frac{1}{\ell^{2}}(2\bar{b}\partial\bar{c}-\bar{c}\partial\bar{b}-T_{C}).
\end{eqnarray}
\end{subequations}As before, the total energy-momentum tensor can
be expressed as $T=\ell^{2}(\hat{T}_{+}+\hat{T}_{-})$.

Therefore, we are naturally led to two different models (A and B,
from now on) depending on the realization of the supersymmetry algebra,
$\{q_{\alpha},q_{\beta}\}=-\gamma_{\alpha\beta}^{m}\partial_{m}$,
with\begin{subequations}
\begin{eqnarray}
q_{\alpha}^{A} & = & \oint\{p_{\alpha}+\frac{1}{2}P_{m}(\gamma^{m}\theta)_{\alpha}-\frac{1}{2\ell^{2}}\partial X^{m}(\gamma_{m}\theta)_{\alpha}-\frac{1}{12\ell^{2}}(\theta\gamma^{m}\partial\theta)(\gamma_{m}\theta)_{\alpha}\},\\
q_{\alpha}^{B} & = & \oint\{p_{\alpha}+\frac{1}{2}P_{m}(\gamma^{m}\theta)_{\alpha}+\frac{1}{2\ell^{2}}\partial X^{m}(\gamma_{m}\theta)_{\alpha}+\frac{1}{12\ell^{2}}(\theta\gamma^{m}\partial\theta)(\gamma_{m}\theta)_{\alpha}\},
\end{eqnarray}
\end{subequations} and the heterotic sectorized model is two-fold,
resembling the two possible descriptions of the heterotic chiral string
studied in \cite{Huang:2016bdd}\footnote{Note that the only other possible realization of the supersymmetry
generator is $q_{\alpha}=\oint\{p_{\alpha}+\frac{1}{2}P_{m}(\gamma^{m}\theta)_{\alpha}\}$
, which corresponds to the ambitwistor case and to which $q_{\alpha}^{A}$
and $q_{\alpha}^{B}$ converge in the $\ell\to\infty$ limit.}. The A model was discussed in section \ref{sec:Amodel} and its spectrum
contains only massless states. The B model will be defined by the
BRST charge\begin{subequations}
\begin{eqnarray}
\hat{Q} & = & \hat{Q}_{\lambda}+Q_{-},\label{eq:BRSTscaleB}\\
\hat{Q}_{\lambda} & \equiv & \oint\,\lambda^{\alpha}\hat{d}_{\alpha},\\
Q_{-} & \equiv & \oint\,\bar{c}(\hat{T}_{-}+\frac{1}{\ell^{2}}\bar{b}\partial\bar{c}),
\end{eqnarray}
\end{subequations}\emph{cf}. equations \eqref{eq:susyinvariantsB}
and \eqref{eq:T+T-het-ell-massive}. Nilpotency of $\hat{Q}$ is straightforward
to demonstrate and its physical content will be discussed next. 

\subsection{Cohomology}

The massless cohomology in the B model is easily obtained, having
the same physical content of the A model. The ghost number two vertex
operators are given by
\begin{eqnarray}
U_{SYM} & = & \lambda^{\alpha}\bar{c}A_{\alpha}^{I}J_{I},\\
U_{SG} & = & \lambda^{\alpha}\bar{c}A_{\alpha}^{m}\hat{P}_{m}^{-},\label{eq:AMBIsugracovariant}
\end{eqnarray}
with equations of motion and gauge transformations analogous to the
ones displayed in \eqref{eq:SYMvertex} and \eqref{eq:SUGRAvertex}.

Now we will consider a vertex operator of the form $U=\bar{c}U_{\textrm{open}}$,
with
\begin{eqnarray}
U_{\textrm{open}} & \equiv & \lambda^{\alpha}\partial\theta^{\beta}B_{\alpha\beta}+(\hat{d}_{\beta},\lambda^{\alpha}C_{\alpha}^{\beta})+(J,\lambda^{\alpha}E_{\alpha})\nonumber \\
 &  & +(N^{mn},\lambda^{\alpha}F_{\alpha mn})+\partial\lambda^{\alpha}G_{\alpha}+(\hat{P}_{m}^{+},\lambda^{\alpha}H_{\alpha}^{m}).\label{eq:massive-open}
\end{eqnarray}
Here, $J=-w_{\alpha}\lambda^{\alpha}$ is the ghost number current
associated to the pure spinor conjugate pair and $N^{mn}=-\frac{1}{2}(w\gamma^{mn}\lambda)$
is the ghost Lorentz contribution. The superfields $B_{\alpha\beta}$,
$C_{\alpha}^{\beta}$, $E_{\alpha}$, $F_{\alpha mn}$, $G_{\alpha}$
and $H_{\alpha}^{m}$ were conveniently chosen to resemble the covariant
description of the first massive level of the pure spinor open string
in \cite{Berkovits:2002qx}. Due to the pure spinor constraint, not
all superfields in $U_{\textrm{open}}$ are independent, for it implies
that
\begin{equation}
(N^{mn},\lambda^{\alpha})\gamma_{\alpha\beta}^{p}\eta_{np}+\frac{1}{2}(J,\lambda^{\alpha})\gamma_{\alpha\beta}^{m}=-2\gamma_{\alpha\beta}^{m}\partial\lambda^{\alpha}.\label{eq:PSproperty1}
\end{equation}
Therefore, $U_{\textrm{open}}$ is invariant under the transformations
parametrized by an arbitrary superfield $\Omega_{m}^{\alpha}$,\begin{subequations}\label{eq:openinvariance}
\begin{eqnarray}
\delta E_{\alpha} & = & \gamma_{\alpha\beta}^{m}\Omega_{m}^{\beta},\\
\delta F_{\alpha mn} & = & \eta_{np}\gamma_{\alpha\beta}^{p}\Omega_{m}^{\beta}-\eta_{mp}\gamma_{\alpha\beta}^{p}\Omega_{n}^{\beta},\\
\delta G_{\alpha} & = & 4\gamma_{\alpha\beta}^{m}\Omega_{m}^{\beta}.
\end{eqnarray}
\end{subequations}

Now we can turn to the the action of the BRST charge on $U$,
\begin{equation}
[\hat{Q},U]=-\bar{c}\{\hat{Q}_{\lambda},U_{\textrm{open}}\}+(\frac{1}{\ell^{2}}-\frac{\Box}{4})\bar{c}\partial\bar{c}U_{\textrm{open}}.\label{eq:BRSTopen}
\end{equation}
As before, BRST-closedness requires that the two terms on the right
hand side above vanish independently. Being careful with the ordering
prescription, one can show that
\begin{eqnarray}
\{\hat{Q}_{\lambda},U_{\textrm{open}}\} & = & \lambda^{\alpha}\lambda^{\gamma}\partial\theta^{\beta}(-D_{\gamma}B_{\alpha\beta}+\frac{2}{\ell^{2}}\gamma_{\beta\gamma}^{m}H_{\alpha}^{n}\eta_{mn})\nonumber \\
 &  & +(\hat{P}_{m}^{+},\lambda^{\alpha}\lambda^{\beta}[D_{\beta}H_{\alpha}^{m}-\gamma_{\beta\gamma}^{m}C_{\alpha}^{\gamma}])\nonumber \\
 &  & -(\hat{d}_{\beta},\lambda^{\alpha}\lambda^{\gamma}[\delta_{\gamma}^{\beta}E_{\alpha}+D_{\gamma}C_{\alpha}^{\beta}+\frac{1}{2}(\gamma^{mn})_{\hphantom{\beta}\gamma}^{\beta}F_{\alpha mn}])\nonumber \\
 &  & +(J,\lambda^{\alpha}\lambda^{\beta}D_{\beta}E_{\alpha})+(N^{mn},\lambda^{\alpha}\lambda^{\beta}D_{\beta}F_{\alpha mn})\nonumber \\
 &  & +\lambda^{\alpha}\partial\lambda^{\beta}(B_{\alpha\beta}+D_{\alpha}G_{\beta}+\gamma_{\beta\gamma}^{m}\partial_{m}C_{\alpha}^{\gamma})\nonumber \\
 &  & -\lambda^{\alpha}\partial\lambda^{\beta}(D_{\beta}E_{\alpha}+\frac{1}{2}(\gamma^{mn})_{\hphantom{\gamma}\beta}^{\gamma}D_{\gamma}F_{\alpha mn}),\label{eq:BRSTUopen}
\end{eqnarray}
and the vanishing of the above expression immediately implies the
superfield equations of motion\begin{subequations}\label{eq:openeom1}
\begin{eqnarray}
\lambda^{\alpha}\lambda^{\beta}(D_{\beta}H_{\alpha}^{m}-\gamma_{\beta\gamma}^{m}C_{\alpha}^{\gamma}) & = & 0,\\
\lambda^{\alpha}\lambda^{\gamma}(D_{\gamma}B_{\alpha\beta}-\frac{2}{\ell^{2}}\gamma_{\beta\gamma}^{m}H_{\alpha}^{n}\eta_{mn}) & = & 0,\\
\lambda^{\alpha}\lambda^{\gamma}(\delta_{\gamma}^{\beta}E_{\alpha}+D_{\gamma}C_{\alpha}^{\beta}+\frac{1}{2}(\gamma^{mn})_{\hphantom{\beta}\gamma}^{\beta}F_{\alpha mn}) & = & 0.
\end{eqnarray}
\end{subequations}Again, not all the terms in \eqref{eq:BRSTUopen}
are independent as can be seen from the identity:
\begin{equation}
(N^{mn},\lambda^{\alpha}\lambda^{\beta})\gamma_{\gamma\beta}^{p}\eta_{np}+\frac{1}{2}(J,\lambda^{\alpha}\lambda^{\beta})\gamma_{\gamma\beta}^{m}=-\frac{5}{2}(\gamma^{m}\partial\lambda)_{\gamma}\lambda^{\alpha}-\frac{1}{2}(\gamma^{mn}\lambda)^{\alpha}(\gamma_{n}\partial\lambda)_{\gamma}.\label{eq:PSproperty2}
\end{equation}
Taking this into account, the vanishing of $\{\hat{Q}_{\lambda},U_{\textrm{open}}\}$
also implies\begin{subequations}\label{eq:openeom2}
\begin{eqnarray}
\lambda^{\alpha}\lambda^{\beta}D_{\beta}E_{\alpha} & = & (\lambda\gamma_{mpqrs}\lambda)K_{n}^{pqrs}\eta^{mn},\\
\lambda^{\alpha}\lambda^{\beta}D_{\beta}F_{\alpha mn} & = & \frac{1}{2}(\lambda\gamma_{npqrs}\lambda)K_{m}^{pqrs}-\frac{1}{2}(\lambda\gamma_{mpqrs}\lambda)K_{n}^{pqrs},
\end{eqnarray}
\begin{multline}
\lambda^{\alpha}\partial\lambda^{\beta}(B_{\alpha\beta}+D_{\alpha}G_{\beta}-D_{\beta}E_{\alpha}-\frac{1}{2}(\gamma^{mn})_{\hphantom{\gamma}\beta}^{\gamma}D_{\gamma}F_{\alpha mn}+\gamma_{\beta\gamma}^{m}\partial_{m}C_{\alpha}^{\gamma})=\\
=-2\lambda^{\alpha}\partial\lambda^{\beta}(\gamma^{m}\gamma_{npqr})_{\beta\alpha}K_{m}^{npqr}+8\lambda^{\alpha}\partial\lambda^{\beta}(\gamma_{npq})_{\beta\alpha}K_{m}^{npqm},
\end{multline}
\end{subequations}where $K_{m}^{npqr}$ is an arbitrary superfield. 

Concerning BRST-exact contributions, $U_{\textrm{open}}$ is defined
up to the following transformations\begin{subequations}\label{eq:opengauge}
\begin{eqnarray}
\delta B_{\alpha\beta} & = & D_{\alpha}\Sigma_{\beta}+\frac{2}{\ell^{2}}\gamma_{\alpha\beta}^{n}\Sigma^{m}\eta_{mn},\\
\delta C_{\alpha}^{\beta} & = & \delta_{\alpha}^{\beta}\Sigma+\frac{1}{2}(\gamma^{mn})_{\hphantom{\beta}\alpha}^{\beta}\Sigma_{mn}+D_{\alpha}\Sigma^{\beta},\\
\delta E_{\alpha} & = & -D_{\alpha}\Sigma,\\
\delta F_{\alpha mn} & = & -D_{\alpha}\Sigma_{mn},\\
\delta G_{\alpha} & = & D_{\alpha}\Sigma-\Sigma_{\alpha}-\gamma_{\alpha\beta}^{m}\partial_{m}\Sigma^{\beta}+\frac{1}{2}(\gamma^{mn})_{\hphantom{\beta}\alpha}^{\beta}D_{\beta}\Sigma_{mn},\\
\delta H_{\alpha}^{m} & = & \gamma_{\alpha\beta}^{m}\Sigma^{\beta}+D_{\alpha}\Sigma^{m}.
\end{eqnarray}
\end{subequations}Here, $\Sigma$, $\Sigma_{\alpha}$, $\Sigma^{\alpha}$,
$\Sigma^{m}$ and $\Sigma_{mn}$ are superfield parameters. It is
not difficult to show that the superfield equations of motion in \eqref{eq:openeom1}
and \eqref{eq:openeom2} are invariant under \eqref{eq:opengauge}.

Up to irrelevant constants, the very same equations of motion were
obtained in \cite{Berkovits:2002qx} in a different context, corresponding
to a superfield description of the first massive level of the open
string. In our case, the mass is given in terms of the dimensionful
parameter $\ell$, such that $M^{2}=\frac{4}{\ell^{2}}$. This will
be proven below, precisely matching the remaining condition for the
vertex $U$ to be BRST-closed, \emph{cf}. equation \eqref{eq:BRSTopen}.
Note that it is not possible to accommodate any other massive states
in the cohomology of the heterotic B model and such massive level
only exists because of a simple interplay between the two sectors. 

The presence of an open string like state, \emph{i.e.} a state matching
the quantum numbers of the first massive level of the open string,
is characteristic to the sectorized models. Next, we will further
explore this feature, discussing the field content of the massive
cohomology and determining the physical degrees of freedom in order
to explicitly write their vertex operator in the light-cone gauge.

\subsection{Light-cone description of the massive states}

The first massive level of the open string consists of a symmetric,
traceless tensor, $G_{mn}$, a 3-form, $A_{mnp}$ and a $\gamma$-traceless
vector-spinor, $\psi_{m\alpha}$, satisfying\begin{subequations}\label{eq:openeom}
\begin{eqnarray}
\eta^{mn}G_{mn} & = & 0,\\
\partial^{m}G_{mn} & = & 0,\\
\partial^{m}\psi_{m\alpha} & = & 0,\\
(\gamma^{m}\psi_{m})^{\alpha} & = & 0,\\
\partial^{m}A_{mnp} & = & 0.
\end{eqnarray}
\end{subequations}These equations are invariant under the $\mathcal{N}=1$
supersymmetry transformations\begin{subequations}
\begin{eqnarray}
\delta G_{mn} & = & M(\epsilon\gamma_{m}\chi_{n})+M(\epsilon\gamma_{n}\chi_{m})-(\epsilon\partial_{m}\psi_{n})-(\epsilon\partial_{n}\psi_{m})\\
\delta\psi_{m\alpha} & = & \frac{1}{4}G_{mn}(\gamma^{n}\epsilon)_{\alpha}+\frac{1}{4}\partial_{q}A_{mnp}(\gamma^{npq}\epsilon)_{\alpha}-\frac{1}{24}\partial_{r}A_{npq}(\gamma_{m}^{\hphantom{m}npqr}\epsilon)_{\alpha}\\
\delta A_{mnp} & = & \frac{1}{3}[(\epsilon\gamma_{mn}\psi_{p})+(\epsilon\gamma_{np}\psi_{m})+(\epsilon\gamma_{pm}\psi_{n})]\nonumber \\
 &  & -\frac{1}{3M}[(\epsilon\gamma_{m}\partial_{n}\chi_{p})+(\epsilon\gamma_{n}\partial_{p}\chi_{m})+(\epsilon\gamma_{p}\partial_{m}\chi_{n})]\nonumber \\
 &  & +\frac{1}{3M}[(\epsilon\gamma_{n}\partial_{m}\chi_{p})+(\epsilon\gamma_{p}\partial_{n}\chi_{m})+(\epsilon\gamma_{m}\partial_{p}\chi_{n})],
\end{eqnarray}
\end{subequations}where $\epsilon^{\alpha}$ is a constant fermionic
parameter, $M$ is the mass and $\chi_{m}^{\alpha}\equiv\frac{1}{M}\gamma_{n}^{\alpha\beta}\partial^{n}\psi_{m\beta}$. 

The supersymmetry algebra closes on-shell,\begin{subequations}
\begin{eqnarray}
[\delta_{1},\delta_{2}]G_{mn} & = & (\epsilon_{1}\gamma^{p}\epsilon_{2})\partial_{p}G_{mn},\\{}
[\delta_{1},\delta_{2}]\psi_{m\alpha} & = & (\epsilon_{1}\gamma^{n}\epsilon_{2})\partial_{n}\psi_{m\alpha},\\{}
[\delta_{1},\delta_{2}]A_{mnp} & = & (\epsilon_{1}\gamma^{q}\epsilon_{2})\partial_{q}A_{mnp},
\end{eqnarray}
\end{subequations}and the matching of the on-shell degrees of freedom
(d.o.f.) can be readily verified. There are 128 bosonic d.o.f. (44
from $G^{mn}$ and $84$ from $A_{mnp}$) and 128 fermionic d.o.f.
coming from $\psi_{m\alpha}$. It is possible of course to extract
all this information about the physical spectrum from the covariant
superfield formulation \cite{Berkovits:2002qx}, but the demonstration
is lengthy and purposeless for our present goal. 

Alternatively, we will take advantage of the DDF-like construction
of \cite{Jusinskas:2014vqa} and propose an $SO(8)$ covariant version
of $U_{\textrm{open}}$ given in \eqref{eq:massive-open}. This will
allow us to pinpoint the physical polarizations and to have a better
knowledge of their behaviour in the $\ell\to\infty$ limit that will
be analyzed in the next section. In order to do it properly, we will
need a quick review of the $SO(8)$ notation. Spacetime vectors, $v^{m}$,
will be decomposed in transversal components, $v^{i}$, with $i=1,\ldots,8$,
and light-cone components, $\sqrt{2}v^{\pm}=(v^{0}\pm v^{9})$. Spacetime
spinors, $s^{\alpha}$ ($t_{\alpha}$), will be written in terms of
the two $SO(8)$ chiralities, $s^{a}$ and $\bar{s}^{\dot{a}}$
($t_{a}$ and $\bar{t}_{\dot{a}}$), with $a,\dot{a}=1,\ldots,8$.
To keep the notation simple, we will use only diagonal $SO(8)$ metrics,
$\{\eta_{ij},\eta_{ab},\eta_{\dot{a}\dot{b}}\}$, and their inverse,
$\{\eta^{ij},\eta^{ab},\eta^{\dot{a}\dot{b}}\}$, making no distinction
between upper and lower indices. The matrices $\gamma^{m}$ will be
conveniently written in terms of the $8$-dimensional equivalent of
the Pauli matrices, $\sigma_{a\dot{a}}^{i}$, with non-vanishing components
given by\label{eq:gammadecomposition}
\begin{equation}
\begin{array}{rclrcl}
\gamma_{\alpha\beta}^{i} & = & \sigma_{a\dot{a}}^{i}, & (\gamma^{i})^{\alpha\beta} & = & \sigma_{b\dot{b}}^{i}\eta^{ab}\eta^{\dot{a}\dot{b}},\\
\gamma_{\alpha\beta}^{+} & = & -\sqrt{2}\eta_{ab}, & (\gamma^{+})^{\alpha\beta} & = & \sqrt{2}\eta^{\dot{a}\dot{b}},\\
\gamma_{\alpha\beta}^{-} & = & -\sqrt{2}\eta_{\dot{a}\dot{b}}, & (\gamma^{-})^{\alpha\beta} & = & \sqrt{2}\eta^{ab},
\end{array}
\end{equation}
Note that the $\sigma^{i}$ matrices satisfy\begin{subequations}\label{eq:pauliproperties}
\begin{eqnarray}
(\sigma_{a\dot{a}}^{i}\sigma_{b\dot{b}}^{j}+\sigma_{a\dot{b}}^{i}\sigma_{b\dot{a}}^{j})\eta_{ij} & = & 2\eta_{ab}\eta_{\dot{a}\dot{b}},\\
(\sigma_{a\dot{a}}^{i}\sigma_{b\dot{b}}^{j}+\sigma_{a\dot{b}}^{i}\sigma_{b\dot{a}}^{j})\eta^{\dot{a}\dot{b}} & = & 2\eta^{ij}\eta_{ab},\\
(\sigma_{a\dot{a}}^{i}\sigma_{b\dot{b}}^{j}+\sigma_{a\dot{b}}^{i}\sigma_{b\dot{a}}^{j})\eta^{ab} & = & 2\eta^{ij}\eta_{\dot{a}\dot{b}}.
\end{eqnarray}
\end{subequations}It will be useful to define also $\sigma^{ij}\equiv\frac{1}{2}(\sigma^{i}\sigma^{j}-\sigma^{j}\sigma^{i})$,
which can be viewed as the spinor Lorentz matrices in $SO(8)$.

Next, we define the $SO(8)$-covariant superfield 
\begin{equation}
F_{\dot{a}a}(a,\bar{a},\xi,\bar{\xi},k,p)\equiv\bar{A}_{\dot{a}}(\bar{a},\bar{\xi},k)A_{a}(a,\xi,p),
\end{equation}
with
\begin{eqnarray}
\bar{A}_{\dot{a}}(\bar{a},\bar{\xi},k) & = & \{\eta_{il}-\frac{ik}{3!}\theta_{il}-\frac{k^{2}}{5!}\theta_{ij}\theta_{jl}+\frac{ik^{3}}{7!}\theta_{ij}\theta_{jk}\theta_{kl}\}\bar{a}^{i}(\sigma^{l}\theta)_{\dot{a}}e^{-ik\sqrt{2}X^{+}}+\frac{i}{k}\bar{\xi}_{\dot{a}}e^{-ik\sqrt{2}X^{+}}\nonumber \\
 &  & +\{\frac{1}{2!}\eta_{il}-\frac{ik}{4!}\theta_{il}-\frac{k^{2}}{6!}\theta_{ij}\theta_{jl}+\frac{ik^{3}}{8!}\theta_{ij}\theta_{jk}\theta_{kl}\}(\bar{\xi}\sigma^{i}\theta)(\sigma^{l}\theta)_{\dot{a}}e^{-ik\sqrt{2}X^{+}},\\
A_{a}(a,\xi,p) & = & \{\eta_{il}-\frac{ip}{3!}\bar{\theta}_{il}-\frac{p^{2}}{5!}\bar{\theta}_{ij}\bar{\theta}_{jl}+\frac{ip^{3}}{7!}\bar{\theta}_{ij}\bar{\theta}_{jk}\bar{\theta}_{kl}\}a^{i}(\sigma^{l}\bar{\theta})_{a}e^{-ip\sqrt{2}X^{-}}+\frac{i}{p}\xi_{a}e^{-ip\sqrt{2}X^{-}}\nonumber \\
 &  & +\{\frac{1}{2!}\eta_{il}-\frac{ip}{4!}\bar{\theta}_{il}-\frac{p^{2}}{6!}\bar{\theta}_{ij}\bar{\theta}_{jl}+\frac{ip^{3}}{8!}\bar{\theta}_{ij}\bar{\theta}_{jk}\bar{\theta}_{kl}\}(\xi\sigma^{i}\bar{\theta})(\sigma^{l}\bar{\theta})_{a}e^{-ip\sqrt{2}X^{-}}.
\end{eqnarray}
These objects correspond to the $SO(8)$-covariant super Maxwell fields
of \cite{Brink:1983pf}, with $(\bar{a}_{i},\bar{\xi}_{\dot{a}})$
and $(a_{i},\xi_{a})$ denoting the polarization pairs for the vector
and the spinor in each frame. Note that $\bar{A}_{\dot{a}}$
($A_{a}$) depends on $\theta_{a}$ ($\bar{\theta}_{\dot{a}}$)
only through $\theta^{ij}=\theta_{a}\sigma_{ab}^{ij}\theta_{b}$ ($\bar{\theta}^{ij}=\bar{\theta}_{\dot{a}}\sigma_{\dot{a}\dot{b}}^{ij}\bar{\theta}_{\dot{b}}$
). The vector superfields are defined by\begin{subequations}
\begin{eqnarray}
D_{a}\bar{A}_{\dot{a}}(k) & = & \sigma_{a\dot{a}}^{i}\bar{A}_{i}(k),\\
\bar{D}_{\dot{a}}A_{a}(p) & = & \sigma_{a\dot{a}}^{i}A_{i}(p),
\end{eqnarray}
\end{subequations}and can be used to show that $F_{a\dot{a}}$ defines
a superfield set $\{F_{a\dot{a}},F_{ia},F_{\dot{a}i},F_{ij}\}$ satisfying
\begin{equation}
\begin{array}{rclcrcl}
D_{b}F_{\dot{a}a} & = & \sigma_{b\dot{a}}^{i}F_{ia}, &  & D_{a}F_{\dot{a}i} & = & \sigma_{a\dot{a}}^{j}F_{ji},\\
\bar{D}_{\dot{b}}F_{\dot{a}a} & = & -\sigma_{\dot{b}a}^{i}F_{\dot{a}i}, &  & \bar{D}_{\dot{b}}F_{\dot{a}i} & = & -ip\sigma_{\dot{b}a}^{i}F_{\dot{a}a},\\
D_{b}F_{ia} & = & ik\sigma_{b\dot{a}}^{i}F_{\dot{a}a}, &  & D_{a}F_{ij} & = & ik\sigma_{a\dot{a}}^{i}F_{\dot{a}j},\\
\bar{D}_{\dot{a}}F_{ia} & = & \sigma_{a\dot{a}}^{j}F_{ij}, &  & \bar{D}_{\dot{a}}F_{ij} & = & ip\sigma_{a\dot{a}}^{j}F_{ia},
\end{array}\label{eq:superF}
\end{equation}
which will be the fundamental blocks in the construction of the $SO(8)$-covariant
form of $U_{\textrm{open}}$. To build it explicitly, we can just
go through the DDF-like procedure described in \cite{Jusinskas:2014vqa}
for generating the massive vertices and adapt it to the sectorized
model. The result is
\begin{eqnarray}
U_{\textrm{open}}(a,\bar{a},\xi,\bar{\xi},k,p) & = & (\hat{P}_{i}^{+},\bar{\lambda}_{\dot{a}}F_{\dot{a}i})-\frac{1}{\sqrt{2}}\sigma_{a\dot{a}}^{i}(\hat{P}_{+}^{+},\lambda_{a}F_{\dot{a}i})-ip\sqrt{2}(N_{i},\bar{\lambda}_{\dot{a}}F_{\dot{a}i})\nonumber \\
 &  & -\frac{2}{\ell^{2}}\bar{\lambda}_{\dot{a}}\partial\theta_{a}F_{\dot{a}a}+ip(\bar{\lambda}\sigma^{j}\sigma^{i}\partial\bar{\theta})F_{ij}-ip(\hat{d}_{a},\bar{\lambda}_{\dot{a}}F_{\dot{a}a}).\label{eq:LC-massive-open}
\end{eqnarray}
Here, $\hat{P}_{i}^{+}$ and $\hat{P}_{+}^{+}$ are just components
of $\hat{P}_{m}^{+}$ and $N^{i}\equiv N^{-i}$. The above expression
corresponds to the vertex depicted in \eqref{eq:massive-open} in
the light-cone frame, after the removal of auxiliary superfields through
\eqref{eq:openinvariance} and \eqref{eq:opengauge}. It is now simple
to determine the mass shell, since
\begin{eqnarray}
\{\hat{Q}_{\lambda},U_{\textrm{open}}\} & = & 2(kp-\frac{1}{\ell^{2}})(\bar{\lambda}\partial\bar{\theta})\lambda^{a}\sigma_{a\dot{a}}^{i}F_{\dot{a}i}\nonumber \\
 &  & -2(kp-\frac{1}{\ell^{2}})(\lambda\sigma^{i}\partial\bar{\theta})\bar{\lambda}_{\dot{a}}F_{\dot{a}i}\nonumber \\
 &  & +2(kp-\frac{1}{\ell^{2}})\bar{\lambda}_{\dot{a}}\partial\lambda_{a}F_{\dot{a}a},\label{eq:BRSTclosednessOPEN}
\end{eqnarray}
which vanishes only for $kp=\frac{1}{\ell^{2}}$, \emph{i.e.} $M^{2}=\frac{4}{\ell^{2}}$.
Therefore, the vertex $U=\bar{c}U_{\textrm{open}}$ describes
physical states in the massive cohomology of the B model. 

The physical degrees of freedom are easily readable from \eqref{eq:LC-massive-open}.
Let us extract the vertex operators from each polarization set by
defining
\begin{equation}
\begin{array}{rclcrcl}
U_{j,i} & \equiv & \frac{\partial}{\partial\bar{a}_{i}}\frac{\partial}{\partial a^{j}}U, &  & U_{a,i} & \equiv & \frac{\partial}{\partial\bar{a}_{i}}\frac{\partial}{\partial\xi^{a}}U,\\
U_{j,\dot{a}} & \equiv & \frac{\partial}{\partial\bar{\xi}_{\dot{a}}}\frac{\partial}{\partial a^{j}}U, &  & U_{a,\dot{a}} & \equiv & \frac{\partial}{\partial\bar{\xi}_{\dot{a}}}\frac{\partial}{\partial\xi^{a}}U.
\end{array}\label{eq:SO8multiplet-massive}
\end{equation}
The symmetric part of $U_{j,i}$ and the vector $\sigma_{a\dot{a}}^{i}U_{a,\dot{a}}$
combine to give an $SO(9)$ symmetric traceless tensor of rank 2,
with 44 d.o.f.. The antisymmetric part of $U_{j,i}$ and the the 3-form
$\sigma_{a\dot{a}}^{ijk}U_{a,\dot{a}}$ combine into an $SO(9)$ 3-form
($\sigma^{ijk}$ is just the antisymmetrized product of three $\sigma$
matrices), with 84 d.o.f.. As for the 128 fermionic degrees of freedom,
$U_{a,i}$ and $U_{j,\dot{a}}$ constitute a $\gamma$-traceless $SO(9)$
vector-spinor. These are the irreducible massive representations of
the fields $G_{mn}$, $A_{mnp}$ and $\psi_{m\alpha}$, respectively,
with equations of motion given in \ref{eq:openeom}. This finishes
the demonstration that the heterotic B model in the sectorized string
contains a massive open string like state.

In the next section we will analyze how the sectorized model is connected
to the ambitwistor string, showing what happens to the massive degrees
of freedom in the infinite length limit.

\section{The $\ell\to\infty$ limit and the connection to ambitwistor strings\label{sec:ambitwistor}}

In the ambitwistor string debutant paper \cite{Mason:2013sva}, Mason
and Skinner pointed out that the model remarkably shares features
of both $\alpha'\to0$ (infinite tension string) and $\alpha'\to\infty$
(null string) limits. This was further explored in \cite{Casali:2016atr}
to determine a connection between the ambitwistor string and the null
string. In this section we will give an explicit construction of the
$\ell\to\infty$ limit, showing that the ambitwistor string can be
viewed as a zero tension limit of the sectorized model. Although we
are focusing on the heterotic case, this construction is easily generalized
to the type II string and the bosonic string.

\subsection{The heterotic ambitwistor string in the pure spinor formalism}

By taking the limit $\ell\to\infty$ in the B model described in the
previous section, it is easy to see that the BRST charge \eqref{eq:BRSTscaleB}
reduces to
\begin{equation}
Q=\oint\{\lambda^{\alpha}d_{\alpha}+\frac{1}{4}\bar{c}P^{m}P_{m}\},\label{eq:BRSTambi}
\end{equation}
now with $d_{\alpha}=p_{\alpha}-\frac{1}{2}(\gamma^{m}\theta)_{\alpha}P_{m}$.
In the A model, the only difference is a minus sign in front of the
$P^{2}$ term. $Q$ should be regarded as the BRST charge for the
heterotic ambitwistor string in the pure spinor formalism, in which
the constraint associated to the $(\bar{b},\bar{c})$ ghost
pair is the particle-like Hamiltonian, $H=-\frac{1}{2}P^{2}$. For
comparison, in Berkovits' proposal \cite{Berkovits:2013xba}, $\bar{b}$
and $\bar{c}$ are identified with the (left-moving) Virasoro
ghosts. While both are consistent at the quantum level, the latter
does not completely describe the supergravity states by failing to
express the expected gauge transformations as BRST-exact states.

It is interesting to mention that the difference between \eqref{eq:BRSTambi}
and the BRST charges of the A and B models is the nilpotent symmetry
of the action \eqref{eq:heteroticaction} defined in \cite{Jusinskas:2016qjd}
(which was an extension of the type II construction of \cite{Chandia-Vallilo}),
generated by
\begin{equation}
\mathcal{K}=\frac{1}{\ell^{2}}\oint\{(\lambda\gamma_{m}\theta)\Pi^{m}\}.
\end{equation}
Because of the dimensionful parameter $\ell$, $\mathcal{K}$ cannot
be part of the BRST charge in the infinite length limit\footnote{This might help to understand how to consistently couple the supergravity
background to the ambitwistor string in the pure spinor formulation,
task that so far has only been achieved in the sectorized model.}.

Note also that $T=\ell^{2}(\hat{T}_{+}+\hat{T}_{-})$ and $\mathcal{H}=(\hat{T}_{+}-\hat{T}_{-})$
satisfy\begin{subequations}\label{eq:THalgebra}
\begin{eqnarray}
T(z)\,T(y) & \sim & \frac{2T}{(z-y)^{2}}+\frac{\partial T}{(z-y)},\\
T(z)\,\mathcal{H}(y) & \sim & \frac{2\mathcal{H}}{(z-y)^{2}}+\frac{\partial\mathcal{H}}{(z-y)},\\
\mathcal{H}(z)\mathcal{H}(y) & \sim & \frac{1}{\ell^{2}}\frac{2T}{(z-y)^{2}}+\frac{1}{\ell^{2}}\frac{\partial T}{(z-y)},
\end{eqnarray}
\end{subequations}which in the $\ell\to\infty$ limit corresponds
to the Galilean Conformal Algebra \cite{Bagchi:2009pe} and appears
as the constraint algebra in the tensionless string \cite{Lizzi:1986nv}.
The OPE's in \eqref{eq:THalgebra} are characteristic to the sectorized
strings \cite{Jusinskas:2016qjd} and in fact were later briefly considered
as part of a tensionful model whose tensionless limit gives the ambitwistor
bosonic string \cite{Casali:2016atr}.

Concerning the tree-level amplitudes discussed in section \ref{sec:Amodel},
the introduction of the dimensionful parameter $\ell$ adds a factor
of $\ell^{-2}$ to the $P^{+}P^{+}$ OPE, which ultimately implies
that the Einstein-gravity term $\mathcal{M}_{1}$ appears multiplied
by the same factor. The calculations in the A and B models are essentially
the same and can be schematically cast as
\begin{equation}
\langle U_{1}U_{2}U_{3}\rangle=\pm\frac{1}{\ell^{2}}(\mathrm{\textrm{Einstein}})+(\textrm{Gauss-Bonnet}),
\end{equation}
where the plus sign corresponds to the A model and the minus sign
corresponds to the B model. This is consistent with the results of
\cite{Huang:2016bdd}, where the ambiguity comes from the different
choices of ``flipped'' sector, and is also consistent with the interpretation
of the ambitwistor string as the $\ell\to\infty$ limit of the sectorized
string, since the graviton three-point function in the heterotic ambitwistor
model gives precisely a Gauss\textendash Bonnet term at tree level.

The cohomology of \eqref{eq:BRSTambi} should contain a manifestly
supersymmetric version of the spectrum found in \cite{Mason:2013sva}.
We have of course the super Yang-Mills vertex, $U_{SYM}=\lambda^{\alpha}\bar{c}A_{\alpha}^{I}J_{I}$,
and supergravity vertex, $U_{SG}=\lambda^{\alpha}\bar{c}A_{\alpha}^{m}P_{m}$,
with the expected equations of motion and gauge transformations. In
addition, Mason and Skinner showed that the heterotic ambitwistor
cohomology contains also a 3-form so far of unknown origin. But there
are possibly even more states previously unaccounted for. Maybe the
easiest way to see this is determining the zero-momentum cohomology
of \eqref{eq:BRSTambi}. In particular, we can focus on the following
vertices,\begin{subequations}
\begin{eqnarray}
U_{m} & \equiv & \bar{c}(\lambda\gamma^{n}\theta)(d\gamma_{mn}\theta)+\bar{c}(\lambda\gamma_{m}\theta)J-\frac{3}{4}\bar{c}P^{n}(\lambda\gamma^{p}\theta)(\theta\gamma_{mnp}\theta),\\
U_{mnp} & \equiv & \bar{c}(\lambda\gamma_{[m}\theta)(d\gamma_{np]}\theta)-2\bar{c}(\lambda\gamma_{[m}\theta)N_{np]}\nonumber \\
 &  & -\frac{1}{2}\bar{c}P_{[m}(\theta\gamma_{np]q}\theta)(\lambda\gamma^{q}\theta)+\frac{1}{2}\bar{c}P^{q}(\lambda\gamma_{[m}\theta)(\theta\gamma_{np]q}\theta),
\end{eqnarray}
\end{subequations}which are BRST-closed and cannot be written as
BRST-exact operators.

As we will see next, such states can be interpreted as a massless
limit of the first massive level of the open string,\emph{ }following
the discussion in section \ref{sec:Bmodel}.

\subsection{The light-cone analysis of the ambitwistor cohomology}

If we consider a general vertex operator of the form
\[
U_{\infty}=\lim_{\ell\to\infty}\bar{c}U_{\textrm{open}},
\]
with $U_{\textrm{open}}$ given in \eqref{eq:massive-open}, the equations
of motion can be easily extracted from \eqref{eq:openeom1} and \eqref{eq:openeom2}.
So it seems that the supergravity vertex operator presented above
is only part of a more general solution. To obtain such solution from
scratch in its covariant form (including all the auxiliary superfields
and associated gauge transformations) is again a cumbersome procedure.
What we will do instead, is to solve the mass-shell constraint for
the light-cone vertex \eqref{eq:LC-massive-open} and then take the
infinite length limit. Then, after identifying the relevant degrees
of freedom, we will propose a covariantized version of the results.

Starting with \eqref{eq:massive-open} for $p=\frac{1}{k\ell^{2}}$
(BRST-closedness condition, \emph{cf.} equation \eqref{eq:BRSTclosednessOPEN}),
we can straightforwardly take the $\ell\to\infty$ limit and express
the result as
\begin{equation}
\lim_{\ell\to\infty}\bar{c}U_{\textrm{open}}(a,\bar{a},\xi,\bar{\xi},k)=a^{j}U_{j}(\bar{a},\bar{\xi},k)-\xi^{a}U_{a}(\bar{a},\bar{\xi},k),
\end{equation}
where\begin{subequations}\label{eq:U_infinity}
\begin{eqnarray}
U_{j}(\bar{a},\bar{\xi},k) & = & (P_{j},\bar{c}\bar{\lambda}_{\dot{a}}\bar{A}_{\dot{a}})-\frac{i}{k\sqrt{2}}\{Q,(P_{+},\bar{c}\bar{A}_{j})\},\label{eq:ambiSUGRA}\\
U_{a}(\bar{a},\bar{\xi},k) & = & (P_{i},\bar{c}\bar{\lambda}_{\dot{a}}\bar{A}_{\dot{a}}(\sigma^{i}\bar{\theta})_{a})-\frac{1}{\sqrt{2}}(P_{+},\bar{c}(\lambda\sigma^{i}\bar{A})(\sigma^{i}\bar{\theta})_{a})\nonumber \\
 &  & +(d_{a},\bar{c}\bar{\lambda}_{\dot{a}}\bar{A}_{\dot{a}})+2ik\bar{c}\partial\theta_{a}\bar{\lambda}_{\dot{a}}\bar{A}_{\dot{a}}.\label{eq:AMBI3form}
\end{eqnarray}
\end{subequations}

At first glance, it is remarkable that $U_{\infty}$ can be so simply
expressed in terms of of the super Yang-Mills field $\bar{A}_{\dot{a}}$.
This is a consequence of the DDF construction and does not seem to
survive in the covariant formulation. 

The vertex $U_{j}$, equation \eqref{eq:ambiSUGRA}, corresponds to
a light-cone version of \eqref{eq:AMBIsugracovariant} and encodes
the $SO(8)$-covariant $\mathcal{N}=1$ heterotic supergravity multiplet
up to a BRST exact term. By factoring out the polarizations, $U_{j}$
can be written as
\begin{equation}
U_{j}(\bar{a},\bar{\xi},k)=\bar{a}_{i}U_{j,i}(k)+\bar{\xi}_{\dot{a}}U_{j,\dot{a}}(k).\label{eq:SO8multiplet-sugra}
\end{equation}
The bosonic degrees of freedom (graviton, Kalb-Ramond field and dilaton)
are described by $U_{j,i}$ while $U_{j,\dot{a}}$ is associated to
their superpartners (gravitino and dilatino).

As a feature of the DDF construction, the vertex $U_{a}$ has the
wrong Lorentz spin. This can be checked, for example, by analyzing
the eigenvalue with respect to the Lorentz generator $M^{+-}$ of
the lowest component in the vector polarization. A similar issue was
discussed in \cite{Jusinskas:2015eza} in the context of the DDF-like
description of antifields in the pure spinor formalism. It can be
corrected by rescaling one of the polarizations by the momentum factor
$k$, such that the correct vertex should be defined in terms of $\bar{A}_{\dot{a}}(\frac{\bar{a}}{k},\bar{\xi},k)$
rather than $\bar{A}_{\dot{a}}(\bar{a},\bar{\xi},k)$. Finally,
the vertex can be cast as
\begin{equation}
U_{a}(\bar{a},\bar{\xi},k)=\bar{a}_{i}U_{a,i}(k)+\bar{\xi}_{\dot{a}}U_{a,\dot{a}}(k),\label{eq:SO8multiplet-3form}
\end{equation}
having a simple supersymmetric structure,\begin{subequations}
\begin{eqnarray}
[q_{b},U_{a,\dot{a}}] & = & k\sigma_{b\dot{a}}^{i}U_{a,i},\\
\{q_{b},U_{a,i}\} & = & -i\sigma_{b\dot{a}}^{i}U_{a,\dot{a}}.
\end{eqnarray}
\end{subequations}The field content of $U_{a}$ can be readily determined
and the $SO(8)$ degrees of freedom are represented by a vector, $A_{i}$,
a 3-form, $A_{ijk}$, a spinor, $\bar{\psi}_{\dot{a}}$, and a sigma-traceless
vector-spinor, $\psi_{a}^{i}$, with the following identification
for the vertex operators:\begin{subequations}\label{eq:SO8fields}
\begin{eqnarray}
A_{i} & \to & \sigma_{a\dot{a}}^{i}U_{a,\dot{a}},\\
A_{ijk} & \to & \sigma_{a\dot{a}}^{ijk}U_{a,\dot{a}},\\
\bar{\psi}_{\dot{a}} & \to & \tfrac{1}{8}\sigma_{a\dot{a}}^{i}U_{a,i},\\
\psi_{a}^{i} & \to & U_{a,i}-\tfrac{1}{8}\sigma_{a\dot{a}}^{i}(\sigma_{\dot{a}b}^{j}U_{b,j}).
\end{eqnarray}
\end{subequations}

\begin{figure}[h]
\begin{tikzpicture}   
\matrix (m) [matrix of math nodes,row sep=2em,column sep=3em,minimum width=2em]
{      U_{j,i} & & U_{a,i} & & U_{j,i} & & U_{a,i} \\  
        & & &\underset{\ell\to \infty}{\longrightarrow}& & & \\   
     U_{j,\dot{a}} & & U_{a,\dot{a}} & & U_{j,\dot{a}} & & U_{a,\dot{a}}\\};
alo
\path[-stealth]     
(m-1-1) edge [<->] node [right] {$q_b$} (m-3-1)
(m-1-1) edge [<->] node [below] {$\bar{q}_{\dot{b}}$} (m-1-3)
(m-1-3) edge [<->] node [left] {$q_b$} (m-3-3)  
(m-3-1) edge [<->] node [above] {$\bar{q}_{\dot{b}}$}(m-3-3)
(m-1-5) edge [<->] node [right] {$q_b$} (m-3-5)
(m-1-5) edge [dashed,<-] node [below] {$\bar{q}_{\dot{b}}$} (m-1-7)
(m-1-7) edge [<->] node [left] {$q_b$} (m-3-7)
(m-3-5) edge [dashed,<-] node [above] {$\bar{q}_{\dot{b}}$} (m-3-7);
\end{tikzpicture}
\centering
\caption[Diagram]{On the left we have the $SO(8)$ degrees of freedom of the massive vertex \eqref{eq:LC-massive-open}. On the right side, after taking the limit $\ell \to \infty$, we are left with two partially independent massless multiplets, connected only by the algebra \eqref{eq:wrongSUSY-massless}.}
\label{fig:Diagram}
\end{figure}
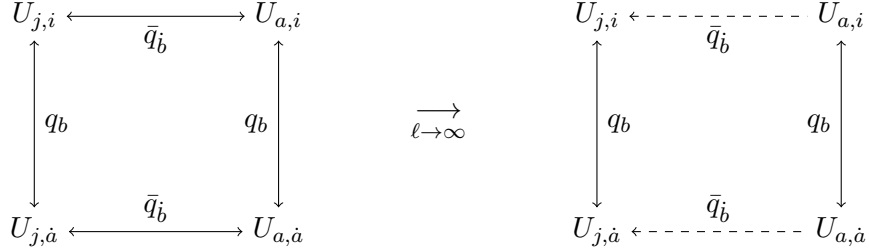

Note that the operators $U_{j}$ and $U_{a}$ are not completely unrelated,
for they satisfy\begin{subequations}\label{eq:wrongSUSY-massless}
\begin{eqnarray}
[\bar{q}_{\dot{a}},U_{a}] & = & -\sigma_{a\dot{a}}^{j}U_{j},\\
\{\bar{q}_{\dot{a}},U_{j}\} & = & 0.
\end{eqnarray}
\end{subequations}Therefore, knowing $U_{a}$ implies knowing $U_{j}$
although the opposite is not true. The above relations also convey
an important information about the BRST triviality of $U_{a}$. Since
the BRST charge commutes with the supersymmetry generators $q_{a}$
and $\bar{q}_{\dot{a}}$, a BRST exact $U_{a}$ is only possible if
the supergravity vertex is also trivial. The light-cone results just
discussed can be conveniently illustrated by the diagram depicted
in Figure \ref{fig:Diagram}, which maps the supersymmetric structure
of the operators $U_{j,i}$, $U_{j,\dot{a}}$, $U_{a,i}$ and $U_{a,\dot{a}}$
before and after the $\ell\to\infty$ limit, \emph{cf.} equations
\eqref{eq:SO8multiplet-massive}, \eqref{eq:SO8multiplet-sugra} and
\eqref{eq:SO8fields}.

In terms of the fields introduced in \eqref{eq:SO8fields}, the following
supersymmetry transformations hold,\begin{subequations}\label{eq:SO8susy}
\begin{eqnarray}
\delta\bar{\psi}_{\dot{a}} & = & \tfrac{3i\sqrt{2}}{64}\partial_{+}A_{i}\sigma_{a\dot{a}}^{i}\epsilon_{a}+\tfrac{i\sqrt{2}}{384}\partial_{+}A_{ijk}\sigma_{a\dot{a}}^{ijk}\epsilon_{a},\\
\delta A_{i} & = & 2i\epsilon_{a}\psi_{a}^{i}-6i\epsilon_{a}\sigma_{a\dot{a}}^{i}\bar{\psi}_{\dot{a}},\\
\delta\psi_{a}^{i} & = & -\tfrac{7i\sqrt{2}}{64}\partial_{+}A_{i}\epsilon_{a}+\tfrac{i\sqrt{2}}{64}\partial_{+}A^{j}\sigma_{ab}^{ij}\epsilon_{b}\nonumber \\
 &  & -\tfrac{5i\sqrt{2}}{128}\partial_{+}A_{ijk}\sigma_{ab}^{jk}\epsilon_{b}+\tfrac{i\sqrt{2}}{128}\partial_{+}A_{jkl}\sigma_{ab}^{ijkl}\epsilon_{b}\\
\delta A_{ijk} & = & 2i\epsilon_{a}\sigma_{a\dot{a}}^{ijk}\bar{\psi}_{\dot{a}}-2i[\epsilon_{a}\sigma_{ab}^{ij}\psi_{b}^{k}+\epsilon_{a}\sigma_{ab}^{jk}\psi_{b}^{i}+\epsilon_{a}\sigma_{ab}^{ki}\psi_{b}^{j}],
\end{eqnarray}
\end{subequations}where $\epsilon_{a}$ is a constant fermionic parameter.

Naturally, we can try to infer a covariant formulation of the dynamics
of these physical degrees of freedom, their equations of motion and
gauge transformations. We found that the bosonic sector can be described
by two vector fields, $A_{m}$ and $B_{m}$, and a 3-form, $A_{mnp}$,
with equations of motion\begin{subequations}\label{eq:covariantEOM}
\begin{eqnarray}
\partial_{m}A^{m} & = & 0,\\
\partial_{m}B^{m} & = & 0,\\
\partial^{p}A_{mnp} & = & \partial_{m}B_{n}-\partial_{n}B_{m}.
\end{eqnarray}
The fermionic sector is represented by a spinor $\psi^{\alpha}$ and
a vector-spinor $\psi_{m\alpha}$ satisfying
\begin{eqnarray}
\partial_{m}(\gamma^{m}\psi)_{\alpha} & = & 0,\\
\partial_{m}(\gamma^{m}\psi_{n})^{\alpha} & = & 0,\\
(\gamma^{m}\psi_{m})^{\alpha} & = & 0,\\
\partial^{m}\psi_{m\alpha} & = & 0.
\end{eqnarray}
\end{subequations}

In addition, we have to supplement the above fields with gauge degrees
of freedom, which can be summarized by the transformations\begin{subequations}\label{eq:covariantGAUGE}
\begin{eqnarray}
\delta A_{m} & = & \partial_{m}\Lambda,\\
\delta\psi_{m\alpha} & = & \partial_{m}\Lambda_{\alpha},\\
\delta B_{m} & = & \partial_{m}\Omega+\partial^{n}\Omega_{mn},\\
\delta A_{mnp} & = & \partial_{m}\Omega_{np}+\partial_{n}\Omega_{pm}+\partial_{p}\Omega_{mn},
\end{eqnarray}
\end{subequations}with gauge parameters $\Lambda$, $\Lambda_{\alpha}$,
$\Omega$ and $\Omega_{mn}$. 

It can be shown that the equations of motion \eqref{eq:covariantEOM}
and gauge transformations \eqref{eq:covariantGAUGE} lead to the light-cone
degrees of freedom discussed before. Note in particular that the field
$B_{m}$ can be gauged to zero but it is required in the covariant
form of the supersymmetry transformations \eqref{eq:SO8susy}, given
by\begin{subequations}
\begin{eqnarray}
\delta\psi^{\alpha} & = & -\frac{3i}{64}\partial_{n}A_{m}(\gamma^{mn}\xi)^{\alpha}-\frac{i}{384}\partial_{q}A_{mnp}(\gamma^{mnpq}\xi)^{\alpha},\\
\delta A_{m} & = & 2i(\xi\psi_{m})-6i(\xi\gamma_{m}\psi),\\
\delta\psi_{m\alpha} & = & \frac{7i}{64}\partial_{n}A_{m}(\gamma^{n}\xi)_{\alpha}+\frac{i}{64}\partial^{n}A^{p}(\gamma_{mnp}\xi)_{\alpha}-\frac{i}{64}\partial_{m}A_{n}(\gamma^{n}\xi)_{\alpha}\nonumber \\
 &  & +\frac{5i}{128}\partial_{q}A_{mnp}(\gamma^{npq}\xi)_{\alpha}-\frac{i}{128}\partial^{r}A^{npq}(\gamma_{mnpqr}\xi)_{\alpha}\nonumber \\
 &  & -\frac{i}{128}\partial_{m}A_{npq}(\gamma^{npq}\xi)_{\alpha}-\frac{i}{32}\partial_{m}B_{n}(\gamma^{n}\xi)_{\alpha},\\
\delta A_{mnp} & = & 2i(\xi\gamma_{mnp}\psi)-2i[(\xi\gamma_{mn}\psi_{p})+(\xi\gamma_{np}\psi_{m})+(\xi\gamma_{pm}\psi_{n})],\\
\delta B_{m} & = & 2i(\xi\gamma_{m}\psi)-2i(\xi\psi_{m}).
\end{eqnarray}
\end{subequations}The supersymmetry algebra closes on-shell up to
gauge transformations,\begin{subequations}
\begin{eqnarray}
[\delta_{1},\delta_{2}]\psi^{\alpha} & = & (\xi_{1}\gamma^{m}\xi_{2})\partial_{m}\psi^{\alpha}\\{}
[\delta_{1},\delta_{2}]A_{m} & = & (\xi_{1}\gamma^{n}\xi_{2})\partial_{n}A_{m}+\textrm{gauge},\\{}
[\delta_{1},\delta_{2}]\psi_{m\alpha} & = & (\xi_{1}\gamma^{n}\xi_{2})\partial_{n}\psi_{m\alpha}+\textrm{gauge},\\{}
[\delta_{1},\delta_{2}]A_{mnp} & = & (\xi_{1}\gamma^{q}\xi_{2})\partial_{q}A_{mnp}+\textrm{gauge},\\{}
[\delta_{1},\delta_{2}]B_{m} & = & (\xi_{1}\gamma^{n}\xi_{2})\partial_{n}B_{m}+\textrm{gauge},
\end{eqnarray}
\end{subequations}

These covariant results might be relevant in trying to demonstrate
the equivalence between the two descriptions of the ambitwistor string,
with worldsheet supersymmetry or using pure spinors. By formulating
a sectorized model of the former, it should be simple to determine
its massive cohomology and verify that it resembles the first level
of the open superstring. If we assume the two descriptions are indeed
equivalent, the cohomology obtained in \cite{Mason:2013sva} for the
heterotic case is incomplete, since its bosonic content is missing
one vector field (here denoted by $A_{m}$). We hope to address this
discrepancy in the future, using the results above as a guiding direction
to a more thorough analysis of the heterotic ambitwistor string using
the RNS formalism. 

In the next section we will summarize our results and open problems,
presenting an overview of the connection between the sectorized and
the ambitwistor strings.

\section{Concluding remarks\label{sec:conclusion}}

\

The sectorized model \cite{Jusinskas:2016qjd} was proposed as a new
interpretation of the modifications introduced by Chand\'ia and Vallilo
\cite{Chandia-Vallilo} to the pure spinor version of the ambitwistor
string \cite{Berkovits:2013xba}. As such, the role of a dimensionful
parameter was never really considered before. In the present work,
we focused on the heterotic case to show that the sectorized model
indeed accommodates a length parameter. That turned out to be a fortunate
choice. Had we decided to analyze the type II case instead, no other
states in the cohomology would have been found. The reason is that
massive states in the sectorized string are open string like\footnote{As pointed out in the text, such states mimic the degrees of freedom
of the first massive level of the open (super)string, hence the name.}, therefore incompatible with $\mathcal{N}=2$ supersymmetry. In other
words, massive states can appear only in the bosonic and heterotic
cases.

We pointed out that the length parameter shares \emph{a priori} no
relation with $\alpha'$. The reason is that the ambitwistor string
was originally built as an infinite tension string ($\alpha'\to0$)
while in the sectorized model it appears as the $\ell\to\infty$ limit.
On the other hand, identifying $\ell^{2}\sim\alpha'$ apparently leads
to no contradiction, as the opposite limits are taken in different
theories.

An interesting property of the sectorized string is made explicit
when we write everything down in terms of the tension $\mathcal{T}\sim\ell^{-2}$
(or $\mathcal{T}\sim\tfrac{1}{\alpha'}$). In this case, the heterotic
models considered here (what we called A and B, \emph{cf}. section
\eqref{sec:Bmodel}, and ambitwistors) can be seen as three different
facets of the same underlying theory, corresponding respectively to
the $\mathcal{T}<0$, $\mathcal{T}>0$ and $\mathcal{T}=0$ regimes.
If one starts with the B model ($\mathcal{T}>0$), the physical spectrum
comprises $\mathcal{N}=1$ supergravity and another multiplet corresponding
to the first massive level of the open superstring. Next, going to
$\mathcal{T}=0$ (ambitwistor string), part of the massive spectrum
collapses to a supergravity multiplet while the remaining states constitute
a new supersymmetric multiplet that includes the 3-form found by Mason
and Skinner \cite{Mason:2013sva}. Moving on to the $\mathcal{T}<0$
region (A model), the supergravity sector remains unchanged and the
BRST closedness conditions insinuate a tachyonic state mirroring the
massive level but forbidden by supersymmetry (note that this is a
fundamental feature of the pure spinor formalism). In the absence
of supersymmetry there should be $M^{2}<0$ states with otherwise
the same quantum numbers as the first massive level of the open string
spectrum. A similar feature appeared already in \cite{Hwang:1998gs},
but such states were disregarded as unphysical. In \cite{Casali:2016atr}
this is further explored in a chiral bosonic model and the cohomology
is partially suggested. The bosonic string seems to be the most adequate
toy model to understand more fundamental aspects of the sectorized
string, including the role of the integrated vertices. This is currently
under investigation and new results should be available soon. 

It is interesting to note that our results share some similarities
with the chiral strings discussed by Huang \emph{et al} in \cite{Huang:2016bdd},
who focused on the factorization of chiral amplitudes. We believe
there are still missing connections between the different chiral strings
and the subject deserves further attention, including its relation
to double field theory and the results of \cite{Hohm:2013jaa}. 

Perhaps the most intriguing (missing) piece concerns a practical construction
of the integrated vertex operator, since the usual construction involving
the $b$ ghost does not render a sensible operator. Recall that for
the pure spinor sectorized string, the energy-momentum tensor, $T$,
and the generalized particle-like Hamiltonian, $\mathcal{H}$, were
shown to be BRST exact in \cite{Jusinskas:2016qjd}, with the definition
of composite operators $b$ and $\tilde{b}$ satisfying
\begin{eqnarray}
\{Q,b\} & = & \frac{1}{\ell^{2}}T,\\
\{Q,\tilde{b}\} & = & \mathcal{H}.
\end{eqnarray}
Curiously, there does not seem to exist an operator that trivializes
$T$ in the $\ell\to\infty$ limit, which might be related to the
absence of the Virasoro constraint in the BRST charge \eqref{eq:BRSTambi}.
Using $b$ and $\tilde{b}$ above, the naive definition of the integrated
vertex operator, $V$, would be
\begin{equation}
V\equiv\int d^{2}z\,\{b_{-1},[\tilde{b}_{-1},U]\},
\end{equation}
where $U$ is the corresponding unintegrated vertex operator and the
subscript $-1$ denotes the Laurent mode of the operators. It is straightforward
to show that
\begin{equation}
[Q,V]=\int d^{2}z\,\{\frac{1}{\ell^{2}}\partial[\tilde{b}_{-1},U]-[b_{-1},[\mathcal{H}_{-1},U]]\}.\label{eq:BRSTint}
\end{equation}
The surface contribution can be discarded but the second term is not
zero, therefore $V$ is not BRST closed. Furthermore, the integrand
in $V$ has the wrong conformal dimensions, $(2,0)$ instead of $(1,1)$.
In principle, $V$ can be corrected by introducing an abstract BRST
closed operator $\bar{\delta}[\mathcal{H}_{-1}]$ with conformal dimension
$(-1,1)$, satisfying
\begin{equation}
\bar{\delta}[\mathcal{H}_{-1}]\cdot[\mathcal{H}_{-1},U]=0.
\end{equation}
Such construction is possible so far only in the $\ell\to\infty$
limit, corresponding to the usual delta function in the ambitwistor
string vertices. Note also that equation \eqref{eq:BRSTint} is consistent
with the absence of surface terms in the BRST closedness analysis
of the integrated vertices proposed in \cite{Berkovits:2013xba}.
Without a consistent definition of the vertex operators, it does not
seem possible to determine from first principles the interactions
of the model. We hope that the reintroduction of the parameter $\ell$
might help clarify this subject, perhaps through an unconventional
approach to higher-point amplitudes.

Finally, it would be very interesting to find an $\mathcal{N}=1$
supergravity action which incorporates also the multiplet of the 3-form.
The first step in this direction would be to understand the background
coupling semi-classically, as we did in \cite{Azevedo:2016zod} for
pure supergravity. We believe that this model is closely related to
the conformal supergravity described in \cite{deRoo:1991at}. It might
be useful also to examine whether the $\mathcal{T}\neq0$ theory can
give us any information on the coupling of the massive multiplet.
In a recent paper \cite{Johansson:2017srf}, Johansson and Nohle introduced
a gauge theory with terms of the form $(DF)^{2}$, which gives rise
to conformal supergravity when combined with SYM via double copy.
Furthermore, they show that there exists a mass deformation of the
form $m^{2}F^{2}$ such that one recovers conformal gravity in the
$m\to0$ limit and Einstein gravity in the $m\to\infty$ limit. In
this sense, their work can be seen as a field-theoretic description
of our results with the identification $m\sim\ell^{-1}$. It would
be interesting to investigate whether this correspondence could lead
to further insights in both theories.

\textbf{Acknowledgments:} We would like to thank Henrik Johansson
for useful discussions. TA acknowledges financial support from the
Knut and Alice Wallenberg Foundation under grant 2015.0083. RLJ would
like to thank the Grant Agency of the Czech Republic for financial
support under the grant P201/12/G028. 

\appendix

\section{Useful OPE's\label{sec:OPE}}

In this appendix we list the OPE's involving the worldsheet fields
in both A and B models of the sectorized heterotic string, as well
as those in the ghost sector, which are common for both models. These
are useful for the cohomology analysis presented in the main text.

\paragraph*{The A model}

The OPE's in the A model are essentially the same as the ones given
in \cite{Jusinskas:2016qjd}, but now we explicitly include the dimensionful
parameter $\ell$. They are given by:
\begin{equation}
\begin{array}{rclcrcl}
d_{\alpha}(z)d_{\beta}(y) & \sim & -\frac{P_{m}^{-}\gamma_{\alpha\beta}^{m}}{(z-y)}, &  & P_{m}^{\pm}(z)P_{n}^{\pm}(y) & \sim & \mp\frac{2}{\ell^{2}}\frac{\eta_{mn}}{(z-y)^{2}},\\
\\
d_{\alpha}(z)P_{m}^{-}(y) & \sim & -\frac{2}{\ell^{2}}\frac{(\gamma_{m}\partial\theta)_{\alpha}}{(z-y)}, &  & d_{\alpha}(z)\Pi^{m}(y) & \sim & \frac{(\gamma_{m}\partial\theta)_{\alpha}}{(z-y)},\\
\\
P_{m}^{-}(z)\Pi^{n}(y) & \sim & -\frac{\delta_{m}^{n}}{(z-y)^{2}}, &  & P_{m}^{+}(z)\Pi^{n}(y) & \sim & -\frac{\delta_{m}^{n}}{(z-y)^{2}}.
\end{array}
\end{equation}

\paragraph*{The B model}

The OPE set for the B model is very similar to the A model and can
be cast as:
\begin{equation}
\begin{array}{rclcrcl}
\hat{d}_{\alpha}(z)\hat{d}_{\beta}(y) & \sim & -\frac{\hat{P}_{m}^{+}\gamma_{\alpha\beta}^{m}}{(z-y)}, &  & \hat{P}_{m}^{\pm}(z)\hat{P}_{n}^{\pm}(y) & \sim & \mp\frac{2}{\ell^{2}}\frac{\eta_{mn}}{(z-y)^{2}},\\
\\
\hat{d}_{\alpha}(z)\hat{P}_{m}^{+}(y) & \sim & \frac{2}{\ell^{2}}\frac{(\gamma_{m}\partial\theta)_{\alpha}}{(z-y)}, &  & \hat{d}_{\alpha}(z)\Pi^{m}(y) & \sim & \frac{(\gamma_{m}\partial\theta)_{\alpha}}{(z-y)},\\
\\
\hat{P}_{m}^{-}(z)\Pi^{n}(y) & \sim & -\frac{\delta_{m}^{n}}{(z-y)^{2}}, &  & \hat{P}_{m}^{+}(z)\Pi^{n}(y) & \sim & -\frac{\delta_{m}^{n}}{(z-y)^{2}}.
\end{array}
\end{equation}

\paragraph*{Ghost sector}

Finally, the relevant OPE's in the ghost sector are given by:
\begin{equation}
\begin{array}{rclcrcl}
\bar{b}(z)\bar{c}(y) & \sim & \frac{1}{(z-y)}, &  & N^{mn}\left(z\right)N^{pq}\left(y\right) & \sim & 6\frac{\eta^{m[p}\eta^{q]n}}{\left(z-y\right)^{2}}+2\frac{\eta^{m[q}N^{p]n}+\eta^{n[p}N^{q]m}}{\left(z-y\right)},\\
\\
J\left(z\right)\lambda^{\alpha}\left(y\right) & \sim & \frac{\lambda^{\alpha}}{\left(z-y\right)}, &  & T\left(z\right)N^{mn}\left(y\right) & \sim & \frac{N^{mn}}{\left(z-y\right)^{2}}+\frac{\partial N^{mn}}{\left(z-y\right)},\\
\\
N^{mn}\left(z\right)\lambda^{\alpha}\left(y\right) & \sim & \frac{1}{2}\frac{\left(\gamma^{mn}\lambda\right)^{\alpha}}{\left(z-y\right)}, &  & T\left(z\right)J\left(y\right) & \sim & \frac{8}{\left(z-y\right)^{3}}+\frac{J}{\left(z-y\right)^{2}}+\frac{\partial J}{\left(z-y\right)},\\
\\
N^{mn}\left(z\right)J\left(y\right) & \sim & \textrm{regular}, &  & J\left(z\right)J\left(y\right) & \sim & -\frac{4}{\left(z-y\right)^{2}}.
\end{array}
\end{equation}

\end{document}